\documentclass{ws-ijmpa}
\usepackage[super,compress]{cite}
\usepackage{graphicx}
\begin{document}
	\markboth{Amee Kakadiya, Zalak Shah and Ajay Kumar Rai}{Mass spectroscopy and decay properties of singly heavy bottom-strange baryons}
	
	%
	\catchline{}{}{}{}{}
	%
	
	\title{Mass Spectra and Decay Properties of Singly Heavy Bottom-Strange Baryons
	}
	
	\author{AMEE KAKADIYA, ZALAK SHAH and AJAY KUMAR RAI
	}
	
	\address{Department of Physics, Sardar Vallabhbhai National Institute of Technology, Ichchhanath, Surat-395007, Gujarat, India
	\footnote{
		ameekakadiya@gmail.com;zalak.physics@gmail.com;
		 raiajayk@gmail.com}}
	
%
	
	\maketitle
	
%
	\begin{abstract}

We enumerated ground and excited state masses of singly heavy bottom-strange baryons using the Hypercentral Constituent Quark Model (hCQM). The screening potential is used with color-Coulomb potential, as the confining potential. We determine the possible $J^P$ values to recently observed excited states like, $\Xi_b(6100)$, $\Xi_b(6227)$, $\Xi_b(6327)$, $\Xi_b(6333)$ and four $\Omega_b(6316,6330,6340,6350)$ states. The Regge trajectories are plotted in $(J, M^2)$ plane to assign the $J^P$ values and quantum numbers to recently observed states. Using the calculated spectroscopic data, properties such as magnetic moment, transition magnetic moment, radiative decay width, and strong decay width are also investigated for all singly heavy bottom-strange baryonic system.

	\end{abstract}
	
	
	\section{Introduction}
	\label{intro}
Understanding the nature of hadrons containing bottom quarks has been accomplished over the last two decades. There have been numerous experimental\cite{PDG,AaijPRD2021,AaijPRL2018,AaijPRL2020,Aaij2021,Sirunyan2021}
 and theoretical studies\cite{Jia2019,Azizi2021,Xiao2020,Jia2021,Moosavi2020,HeLiang2021} on the mass spectra and properties of heavy baryons. Many bottom baryon decay channels have been observed at various experimental facilities, and a large amount of data on heavy baryons has been gathered. The spectroscopy and properties of heavy baryons are vast and useful topics in hadronic physics. The study of heavy flavour baryons containing \textit{b} or \textit{c} quark can play an emergent role to get deep understanding of QCD.
\begin{table}[h]
		\tbl{Mass, width, $J^p$ value and experimental status of the singly heavy bottom-strange baryons from PDG\cite{PDG}.}
		{\begin{tabular}{@{}cccccc@{}} \toprule
					Resonance & Mass (in MeV) & Width (in MeV)& Mean life ($10^{-12}$ sec) & $J^p$ & Status \\  \colrule
	$\Xi_b^0$ & $5791.9\pm 0.5$ & - & $1.477 \pm 0.026 \pm 0.019$ & $\frac{1}{2}^+$ & $***$\\
	$\Xi_b^-$ & $5797.0\pm 0.5$ & - & $1.57 \pm 0.04$ & $\frac{1}{2}^+$ & $***$\\
	$\Xi_b^{'}(5935)^{-}$ & $5935.02\pm 0.05$ & $<0.08$ &-& $\frac{1}{2}^+$ &$***$\\ 
	$\Xi_{b}(5945)^0$ & $5952.3\pm 0.6$ & $0.90\pm 0.18$ &-& $\frac{3}{2}^+$ &$***$\\
	$\Xi_{b}(5955)^-$ & $5955.33\pm 0.13$& $23.96\pm 0.13$ &-& $\frac{3}{2}^+$ &$***$\\
	$\Xi_{b}(6100)^-$ & $6100.3 \pm 0.2$& $<1.9$ &-& $?^?$ &-\\
	$\Xi_{b}(6227)^0$ & $6226.8\pm 1.6$ & $19_{-4}^{+5}$&- & $?^?$ & $***$\\
	$\Xi_{b}(6227)^-$ & $6227.9\pm 0.9$ & $19.9\pm 2.6$ &-& $?^?$ & $***$\\	
	$\Xi_{b}(6327)^-$ & $6327.3_{-0.21}^{+0.23}$& $<2.56$&- & $?^?$ &-\\
	$\Xi_{b}(6333)^-$ & $6333.3_{-0.18}^{+0.17}$& $<1.85$& -& $?^?$ &-\\
	$\Omega_b^-$ & $6046.1\pm 1.7$ & - & $1.65_{-0.16}^{+0.18}$ & $\frac{1}{2}^+$ & $***$\\
	$\Omega_{b}(6316)^-$ & $6315.6\pm 0.6$ & $<4.2$ &-& $?^?$ & $*$\\
	$\Omega_{b}(6330)^-$ & $6330.3\pm 0.6$ & $<4.7$ &-& $?^?$ & $*$\\
	$\Omega_{b}(6340)^-$ & $6339.7\pm 0.6$ & $<1.8$ &-& $?^?$ & $*$\\
	$\Omega_{b}(6350)^-$ & $6349.8\pm 0.6$ & $<3.2$ &-& $?^?$ & $*$\\	
 \botrule
			\end{tabular} \label{ta1}}
	\end{table}	
	The singly heavy bottom-strange baryons belong to two different $SU(3)$ flavor representations: $3 \otimes 3 = 6_{s} \oplus \bar{3}_{A}$. The $\Omega_{b}$ baryon is a part of $SU(3)$ symmetric sextet, while $\Xi_{b}$ doublet is a part of anti-symmetric anti-triplets. 
	
	\begin{center}
		$\Omega_b^-=ssb$,
		$\Xi_{b}^0=\frac{1}{\sqrt{2}}(us-su)b$,
		$\Xi_{b}^-=\frac{1}{\sqrt{2}}(ds-sd)b$
	\end{center}  
 Large samples of bottom baryon decays have been collected by Hadron collider experiments, allowing for increasingly precise mass and lifetime measurements. The Particle Data Group (PDG)\cite{PDG} recently published a list of eleven new baryonic states in the singly heavy bottom family that have yet to be assigned a $J^p$ values such as, $\Lambda_b(6070)^0$, $\Sigma_b(6097)$, $\Xi_b(6100)^-$, $\Xi_b(6227)^-$, $\Xi_b(6227)^0$, $\Xi_b(6327)^0$, $\Xi_b(6333)^0$, $\Omega_b(6315)^-$, $\Omega_b(6330)^-$, $\Omega_b(6340)^-$, and $\Omega_b(6350)^-$.  The list of all known singlly heavy strange-bottom baryons are shown in Table \ref{ta1} with the mass, width, mean life and status.. 
Spectroscopy of single heavy baryons is an important tool for studying the behaviour of light quarks in the presence of heavy quarks. Singly bottom baryons have been studied using a variety of theoretical and phenomenological approaches to the date, such as  the Heavy Quark Symmetry (HQS)  \cite{WangCPC2017,Mutuk2020,Yamaguchi2015}, relativistic quark­-diquark picture\cite{Ebert2011}, QCD sum rule\cite{ZGWang, Mao2015}, relativistic flux tube model \cite{Chen2015} , Regge Phenomology \cite{Juhi,Wei2017}, Holography Inspired Stringy Hadron (HISH) model \cite{Sonnenschein2019}, non-relativistic constituent quark model\cite{Thakkar2017,Shah2018fbs1,Ghalenovia2014,Santopinto2005,Yoshida,Chen2018,Valcarce,Roberts2008},  ,lattice QCD \cite{Padmanath2017} etc.

In this article, we introduce the non-relativistic approach of hypercentral Constituent Quark Model (hCQM) employing the screening potential as confining potential with color-Coulomb potential. Such a model is well established and have been used to determine the properties of heavy flavored baryons previosly (see in Refs. \cite{Gandhi2018,GandhiIJTP2020,AKakadiya,Universe}).
The goal of this research is to look insights into singly heavy bottom-strange baryons and assign a $J^p$ value to experimentally observed states that don't have one, to plot the Regge trajectory (In $J, M^2$ plane) to justify calculated mass spectra, and to investigate their properties.  The structure of this paper is as follows: Starting with the introduction in section-1, the theoretical framework is explained in section-2. The mass spectra of radial and excited states of $\Xi_b$, $\Xi_b^{'}$, and $\Omega_b$ baryons are tabulated along with the Regge trajectories  in the $(J, M^2)$ planes in section-3. The properties of singly heavy bottom-strange baryons are presented in Section \ref{section4}, including magnetic moment, transition magnetic moment, transition decay widths, and strong decay widths. Finally, summary is presented in section-5.
 

 \section{Theoretical framework}
\label{section2}
The Hypercentral Constituent Quark Model has been utilized to describe the inter-quark interaction. The screening potential \cite{Li2009} is incorporated as a confining potential with color-Coulomb potential. The Jacobi coordinates are employed to understand the the dynamics of the constituent quarks inside the baryonic system, which can be expressed as \cite{ShahCPC2016,Shah2016epja,DAE2019,ICC2019},

\begin{equation}
	\vec{\rho}=\frac{\vec{r_1}-\vec{r_2}}{\sqrt2} \hspace{0.5cm} and \hspace{0.5cm} 
	\vec{\lambda}=\frac{\vec{r_1}+\vec{r_2}-2\vec{r_3}} {\sqrt{6}}.
	\label{eqn:1}
\end{equation} 

\noindent The hyper radius x is a collective variable gives a measure of dimension of the three-quark system and the hyper angle $\xi$ reflects its deformation. We define as \cite{Giannini2015}:

\begin{equation}
	x=\sqrt{\rho^2+\lambda^2} 
	\hspace{0.5cm} and \hspace{0.5cm}
	\xi=arctan\left(\frac{\rho}{\lambda}\right)
	\label{eqn:5}
\end{equation}

\noindent The Hamiltonian of the three quark bound system is given by \cite{Giannini2015, Bijkar2000},

\begin{equation} 
	H=\frac{P^2}{2m} + V(x)
	\label{eqn:2}
\end{equation}

$x$ is the six dimensional radial hyper central coordinate of the three body system. The expression of kinetic energy operator for the baryonic system of three quarks in center-of-mass frame is,

\begin{equation} 
	\frac{P{{_x}^2}}{2m}=-\frac{\hbar^2}{2m}(\Delta_{\rho}+\Delta_{\lambda})=-\frac{\hbar^2}{2m}\left(\frac{\partial^2}{\partial x^2}+\frac{5}{x}\frac{\partial}{\partial x}+\frac{L^2(\Omega)}{x^2}\right)
	\label{eqn:6}
\end{equation}

\noindent Here, $P$ is conjugate momentum, $L^2(\Omega)=L^2(\Omega_{\rho}, \Omega_{\lambda}, \xi)$ is the Grand angular operator, which is the six-dimensional generalization of the squared angular momentum operator. $m$ is the reduced mass of the system, which expressed as, $m=\frac{2m_\rho m_\lambda}{m_\rho + m_\lambda}$. where, $m_{\rho}=\frac{2m_1m_2}{m_1+m_2}$ and $m_{\lambda}=\frac{2m_3(m_{1}^{2}+m_{2}^{2}+m_{1}m_{2})}{(m_1+m_2)(m_1+m_2+m_3)}$ . Here,  $m_1$, $m_2$, $m_3$ are the masses of the constituent quarks: $m_u=m_d=0.344 GeV, m_s=0.500 GeV, m_b=4.670GeV$. \\

$V(x)$ is non-relativistic interaction potential inside the baryonic system, which comes in two terms i.e. spin dependent ($V_{SD}$) and spin independent ($V_{SI}$) potential term \cite{Voloshin2008,Bijkar1994,AKakadiya}.
\begin{equation}
	V(x)=V_{SD}(x) + V_{SI}(x)
\end{equation} 
The spin dependent part of potential $V_{SD}(x)$ contains three interaction terms, which are: the spin-spin interaction term $V_{SS}(x)$, the spin-orbit interaction term $V_{\gamma S}(x)$ and tensor term $V_T(x)$ \cite{ShahCPC2016,Shah2016epja},
\begin{equation}
	V_{SD}(x) = V_{SS}(x)(\vec{S_{\rho}} \cdot \vec{S_{\lambda}}) + V_{\gamma S}(x)(\vec{\gamma} \cdot \vec{S}) +V_T(x) \left[ S^2 - \frac{3 (\vec{S} \cdot \vec{x}) (\vec{S} \cdot \vec{x})}{x^2} \right]
	\end{equation}

\noindent where $\vec{S} = \vec{S_\rho} + \vec{S_\lambda}$, are the spin vector associated with the $\vec{\rho}$ and $\vec{\lambda}$ variables respectively (more details can be found in\cite{Thakkar2017}). The Spin-orbit and the tensor term describe the fine structure of the states, while the spin- spin term gives the spin singlet triplet splittings. 

\begin{equation}
V_{SS}(x)= \dfrac{1}{3 m_{\rho} m_{\lambda}} \bigtriangledown^{2} V_{V}
\end{equation}
\begin{equation}
V_{\gamma S} (x) = \dfrac{1}{2 m_{\rho} m_{\lambda}x}  \left(3\dfrac{dV_{V}}{dx} -\dfrac{dV_{S}}{dx} \right)
\end{equation}
\begin{equation}
V_{T}(x)=\dfrac{1}{6 m_{\rho} m_{\lambda}} \left(3\dfrac{d^{2}V_{V}}{dx^{2}} -\dfrac{1}{x}\dfrac{dV_{V}}{dx} \right)
\end{equation}

\noindent The screened potential is incorporated as confining potential with the color-Coulomb potential (spin independent  potential $V_{SI}(x)=V_{conf}(x) + V_{Col}(x)$)\cite{Gandhi2018,GandhiIJTP2020,AKakadiya,Universe}. 

\begin{equation}
	V_{conf}(x)=a\left(\frac{1-e^{-{\mu} x}}{\mu}\right)
	\label{eqn:7}
\end{equation}

\noindent where, $a$ is the string tension and the constant $\mu$(0.07 GeV) is the the screening factor. When $x\ll \frac{1}{\mu}$, the screened potential becomes linear like potential $ax$ and when $x\gg \frac{1}{\mu}$, it will be a constant $\frac{a}{\mu}$. Hence, it is interesting to generate the mass spectra employing screened potential, which gives the lower mass values of excited states than linear potential \cite{Li2009,Wang2019}.
 And the color-Coulomb potential is,
 \begin{equation}
 	V_{Col}(x)= \frac{\tau}{x}
 	\label{eqn:8}
 \end{equation}
 where,$x$ indicates the inter-quark separation, the hyper-Coulomb strength $\tau=-\frac{2}{3}\alpha_s$ ($\frac{2}{3}$ is color factor for baryon), the parameter $\alpha_s$ corresponds to the strong running coupling constant with value 0.7\cite{ICC2019}.  The short-distance part of the static three-quark system, arising from one-gluon exchange within baryon, is of Coulombic shape. Here, we can observe that the strong running coupling constant ($\alpha_s$) becomes smaller as we decrease the distance, the effective potential approaches the lowest order one-gluon exchange potential given as r$\rightarrow$0. So, for short distances, one can use the one gluon exchange potential, taking into account the running coupling constant $\alpha_s$.

\begin{table}
	\tbl{Predicted masses of radial states of $\Xi_{b}$ baryon (in GeV).}
	{\begin{tabular}{@{}ccccccccccc@{}} \toprule
state & Present &  PDG\cite{PDG} &  \cite{Ebert2011} & \cite{Wei2017} & \cite{Roberts2008} & \cite{Karliner2015} &\cite{Yamaguchi2015}\\

 \colrule
$1S$ & 5.796  & 5.797  & 5.803 & 5.793 & 5.806  & 5.795 & 5.806\\ 
$2S$ & 6.208 &&& 6.266  \\ 
$3S$ & 6.533  &&& 6.601 \\ 
$4S$ & 6.825  &&& 6.913  \\ 
$5S$ & 7.094  &&& 7.165 \\ 
$6S$ & 7.347  &&& 7.415 \\ 
\botrule
\end{tabular}\label{ta2}}	
\end{table}

\begin{table}
	\tbl{Predicted masses of radial states of $\Xi_{b}^{'}$ baryon (in GeV).}
	{\begin{tabular}{@{}cccccccccccccc@{}} \toprule
			state & \multicolumn{2}{c}{Present} & \multicolumn{2}{c}{PDG\cite{PDG}}& \multicolumn{2}{c}{\cite{Ebert2011}} & \multicolumn{2}{c}{\cite{HeLiang2021}} & \cite{Wei2017} \\
			
			 &$S=\frac{1}{2}$ & $S=\frac{3}{2}$ & $S=\frac{1}{2}$ & $S=\frac{3}{2}$& $s=\frac{1}{2}$  & $s=\frac{3}{2}$ & $s=\frac{1}{2}$  & $s=\frac{3}{2}$ & \\   \colrule
		$1S$ & 5.935 & 5.958 & 5.935 & 5.955 & 5.936 & 5.963 & & & 5.935 \\
		$2S$ & 6.328 & 6.343 & & & 6.329 & 6.342 & 6.329 &6.342 \\ 
		$3S$ & 6.625 & 6.634 & & & 6.687 & 6.695 \\ 
		$4S$ & 6.902 & 6.907 & & & 6.978 & 6.984 \\ 
		$5S$ & 7.161 & 7.165 & & & 7.229 & 7.234 \\ 
		$6S$ & 7.405 & 7.408  \\  
			\botrule
		\end{tabular}\label{ta3}}	
\end{table}	

\section{Mass spectra and Regge trajectories}
\label{section3}

Masses of all possible states presented in this work, are calculated using Mathematica notebook \cite{Lucha1999}. And from the calculated masses, the $J^p$ value can be assigned to the newly observed states which have no confirmed $J^p$ value.

\begin{table}
	\tbl{Predicted masses of orbital excited states of $\Xi_{b}$ baryon (in GeV).}
	{\begin{tabular}{@{}ccccccccccc@{}} \toprule
	state & $2S+1$ & $J^p$ & Present & Exp.\cite{Aaij2021,Sirunyan2021} & \cite{Ebert2011} & \cite{WangCPC2017} & \cite{Roberts2008} & \cite{Karliner2015} & \cite{Chen2015}& \cite{Juhi}\\  \colrule
	1P & 2 & $\frac{1}{2}^{-}$ & 6.137 && 6.120 & 6.054 & 6.090 & 6.106 & 6.097\\
	&   & $\frac{3}{2}^{-}$ &\textbf{ 6.135 }&\textbf{ 6.100} & 6.130 & 6.040 & 6.093 & 6.115 & 6.106&6.093\\
	& 4 & $\frac{1}{2}^{-}$ & 6.138 && & 6.079\\
	&   & $\frac{3}{2}^{-}$ & 6.136 && \\
	&   & $\frac{5}{2}^{-}$ & 6.133 &&&&&&& 6.240\\
	\hline   
	2P & 2 & $\frac{1}{2}^{-}$ & 6.341 && 6.496\\
	&   & $\frac{3}{2}^{-}$ & 6.339 && 6.502\\
	& 4 & $\frac{1}{2}^{-}$ & 6.342 & \\
	&   & $\frac{3}{2}^{-}$ & 6.340 & \\
	&   & $\frac{5}{2}^{-}$ & 6.338 & \\
	\hline   
	3P & 2 & $\frac{1}{2}^{-}$ & 6.520 && 6.805\\
	&   & $\frac{3}{2}^{-}$ & 6.519 && 6.810\\
	& 4 & $\frac{1}{2}^{-}$ & 6.521 & \\
	&   & $\frac{3}{2}^{-}$ & 6.520 & \\
	&   & $\frac{5}{2}^{-}$ & 6.518 & \\
	\hline      
	4P & 2 & $\frac{1}{2}^{-}$ & 6.679 && 7.068\\
	&   & $\frac{3}{2}^{-}$ & 6.678 && 7.073\\
	& 4 & $\frac{1}{2}^{-}$ & 6.679 & \\
	&   & $\frac{3}{2}^{-}$ & 6.678 & \\
	&   & $\frac{5}{2}^{-}$ & 6.677 & \\
	\hline
		1D & 2 & $\frac{3}{2}^{+}$ & 6.243 &\textbf{6.327}& 6.366 & & & 6.344\\
		&   & $\frac{5}{2}^{+}$ & 6.240 &\textbf{6.333}& 6.373 & & & 6.349&&6.380\\
		& 4 & $\frac{1}{2}^{+}$ & 6.247 & \\
		&   & $\frac{3}{2}^{+}$ & 6.245 & \\
		&   & $\frac{5}{2}^{+}$ & 6.241 & \\   
		&   & $\frac{7}{2}^{+}$ & 6.237 &&&&&&&6.516 \\
		\hline   
		2D & 2 & $\frac{3}{2}^{+}$ & 6.438 && 6.690\\
		&   & $\frac{5}{2}^{+}$ & 6.436 && 6.696\\
		& 4 & $\frac{1}{2}^{+}$ & 6.440 & \\
		&   & $\frac{3}{2}^{+}$ & 6.439 & \\
		&   & $\frac{5}{2}^{+}$ & 6.437 & \\   
		&   & $\frac{7}{2}^{+}$ & 6.434 & \\   
		\hline   
		3D & 2 & $\frac{3}{2}^{+}$ & 6.610  && 6.966\\
		&   & $\frac{5}{2}^{+}$ & 6.608  && 6.970\\
		& 4 & $\frac{1}{2}^{+}$ & 6.611  & \\
		&   & $\frac{3}{2}^{+}$ & 6.610  & \\
		&   & $\frac{5}{2}^{+}$ & 6.609  & \\   
		&   & $\frac{7}{2}^{+}$ & 6.607  & \\
		\hline   
		4D & 2 & $\frac{3}{2}^{+}$ & 6.762 && 7.208\\
		&   & $\frac{5}{2}^{+}$ & 6.761 && 7.212\\
		& 4 & $\frac{1}{2}^{+}$ & 6.763 & \\
		&   & $\frac{3}{2}^{+}$ & 6.763 & \\
		&   & $\frac{5}{2}^{+}$ & 6.762 & \\   
		&   & $\frac{7}{2}^{+}$ & 6.761 & \\
		\hline
			\end{tabular}\label{ta4}}
	{\it Table 4.}	$(${\it Continued}$)$
\end{table}

\begin{table}
\centering
{\begin{tabular}{@{}ccccccccccc@{}} \toprule
state & $2S+1$ & $J^p$ & Present & PDG\cite{PDG}& \cite{Ebert2011} & \cite{WangCPC2017} & \cite{Roberts2008} & \cite{Karliner2015} & \cite{Chen2015}&\cite{Juhi} \\  \colrule
	1F & 2 & $\frac{5}{2}^{-}$ & 6.336& & 6.577 & & & 6.555\\
	&   & $\frac{7}{2}^{-}$ & 6.331 && 6.581 & & & 6.559&&6.654\\
	& 4 & $\frac{3}{2}^{-}$ & 6.341 & \\
	&   & $\frac{5}{2}^{-}$ & 6.337 & \\
	&   & $\frac{7}{2}^{-}$ & 6.333 & \\   
	&   & $\frac{9}{2}^{-}$ & 6.328 &&&&&&& 6.780\\   
	\hline   
	2F & 2 & $\frac{5}{2}^{-}$ & 6.524 && 6.863\\
	&   & $\frac{7}{2}^{-}$ & 6.521 & &6.867\\
	& 4 & $\frac{3}{2}^{-}$ & 6.527 & \\
	&   & $\frac{5}{2}^{-}$ & 6.524 & \\
	&   & $\frac{7}{2}^{-}$ & 6.522 & \\   
	&   & $\frac{9}{2}^{-}$ & 6.519 & \\      		
			\botrule
		\end{tabular}}	
\end{table}

\begin{table}
	\tbl{Predicted masses of orbital excited states of $\Xi_{b}^{'}$ baryon (in GeV).}
	{\begin{tabular}{@{}ccccccccccc@{}} \toprule
		state & $2S+1$ & $J^p$ & Present & PDG\cite{PDG} & \cite{Ebert2011} & \cite{HeLiang2021} & \cite{Wei2017} & \cite{WangCPC2017}& \cite{Jia2021}&\cite{Juhi}\\   \colrule
	1P & 2 & $\frac{1}{2}^{-}$ & 6.235 && 6.233 & 6.233 && 6.142&6.226&6.229\\
	&   & $\frac{3}{2}^{-}$ & \textbf{6.232} & \textbf{6.227} &6.234 & 6.234 &6.215 & 6.093&6243\\
	& 4 & $\frac{1}{2}^{-}$ & 6.237 && 6.227 & 6.227&& 6.176&6235\\
	&   & $\frac{3}{2}^{-}$ & 6.234 && 6.224 & 6.224&&&6.252 \\
	&   & $\frac{5}{2}^{-}$ & 6.229 && 6.226 & 6.226&&&6.262\\
	\hline   
	2P & 2 & $\frac{1}{2}^{-}$ & 6.494& & 6.611\\
	&   & $\frac{3}{2}^{-}$ & 6.492 && 6.605\\
	& 4 & $\frac{1}{2}^{-}$ & 6.495 && 6.604\\
	&   & $\frac{3}{2}^{-}$ & 6.493 && 6.598\\
	&   & $\frac{5}{2}^{-}$ & 6.490 && 6.596\\
	\hline   
	3P & 2 & $\frac{1}{2}^{-}$ & 6.731& & 6.915\\
	&   & $\frac{3}{2}^{-}$ & 6.729 && 6.905\\
	& 4 & $\frac{1}{2}^{-}$ & 6.732 && 6.906\\
	&   & $\frac{3}{2}^{-}$ & 6.730 && 6.900\\
	&   & $\frac{5}{2}^{-}$ & 6.728 && 6.897\\
	\hline      
	4P & 2 & $\frac{1}{2}^{-}$ & 6.949 && 7.174\\
	&   & $\frac{3}{2}^{-}$ & 6.948 && 7.163\\
	& 4 & $\frac{1}{2}^{-}$ & 6.950 && 7.164\\
	&   & $\frac{3}{2}^{-}$ & 6.949 && 7.159\\
	&   & $\frac{5}{2}^{-}$ & 6.947 && 7.156\\
	\hline
		1D & 2 & $\frac{3}{2}^{+}$ & 6.375&&6.459 & 6.459\\
	&   & $\frac{5}{2}^{+}$ & 6.371 && 6.432 & 6.432 & 6.486\\
	& 4 & $\frac{1}{2}^{+}$ & 6.380 && 6.447 & 6.447\\
	&   & $\frac{3}{2}^{+}$ & 6.377 && 6.431 & 6.431\\
	&   & $\frac{5}{2}^{+}$ & 6.373 && 6.420 & 6.420&&&&6.510\\   
	&   & $\frac{7}{2}^{+}$ & 6.368 && 6.414 & 6.414\\
	\hline   
	2D & 2 & $\frac{3}{2}^{+}$ & 6.628 &&6.775 \\
	&   & $\frac{5}{2}^{+}$ & 6.625 && 6.751\\
	& 4 & $\frac{1}{2}^{+}$ & 6.632 && 6.767\\
	&   & $\frac{3}{2}^{+}$ & 6.630 && 6.751\\
	&   & $\frac{5}{2}^{+}$ & 6.626 && 6.740\\   
	&   & $\frac{7}{2}^{+}$ & 6.621 && 6.736\\   
	\hline   
	3D & 2 & $\frac{3}{2}^{+}$ & 6.859  & \\
	&   & $\frac{5}{2}^{+}$ & 6.856  & \\
	& 4 & $\frac{1}{2}^{+}$ & 6.861  & \\
	&   & $\frac{3}{2}^{+}$ & 6.860  & \\
	&   & $\frac{5}{2}^{+}$ & 6.857  & \\   
	&   & $\frac{7}{2}^{+}$ & 6.854  & \\
	\hline   
	4D & 2 & $\frac{3}{2}^{+}$ & 7.070 & \\
	&   & $\frac{5}{2}^{+}$ & 7.069 & \\
	& 4 & $\frac{1}{2}^{+}$ & 7.072 & \\
	&   & $\frac{3}{2}^{+}$ & 7.071 & \\
	&   & $\frac{5}{2}^{+}$ & 7.069 & \\   
	&   & $\frac{7}{2}^{+}$ & 7.067 & \\ 
	\hline
\end{tabular}\label{ta5}}
{\it Table 7.}	$(${\it Continued}$)$
\end{table}

\begin{table}
\centering
{\begin{tabular}{@{}cccccccccc@{}} \toprule
state & $2S+1$ & $J^p$ & Present& \cite{Ebert2011} & \cite{HeLiang2021} & \cite{Wei2017} & \cite{WangCPC2017}&\cite{Juhi}\\   \colrule

	1F & 2 & $\frac{5}{2}^{-}$ & 6.501 & 6.686 \\
	&   & $\frac{7}{2}^{-}$ & 6.495 & 6.640 & & $6.745\pm 0.049$ \\
	& 4 & $\frac{3}{2}^{-}$ & 6.509 & 6.675\\
	&   & $\frac{5}{2}^{-}$ & 6.503 & 6.640\\
	&   & $\frac{7}{2}^{-}$ & 6.497 & 6.619&&&&6.779\\   
	&   & $\frac{9}{2}^{-}$ & 6.489 & 6.610\\   
	\hline   
	2F & 2 & $\frac{5}{2}^{-}$ & 6.747 & \\
	&   & $\frac{7}{2}^{-}$ & 6.744 & \\
	& 4 & $\frac{3}{2}^{-}$ & 6.752 & \\
	&   & $\frac{5}{2}^{-}$ & 6.749 & \\
	&   & $\frac{7}{2}^{-}$ & 6.745 & \\   
	&   & $\frac{9}{2}^{-}$ & 6.740 & \\      		
			\botrule
		\end{tabular}}	
\end{table}

%

\subsection{$\Xi_b$ State}

One $u \slash d$ quark, one $s$ quark, and one bottom quark make up the $\Xi_b$ baryons. The $\Xi_b$ baryon exists in three states: $\Xi_b^0 (usb)$, $\Xi_b^- (dsb)$, and $\Xi_b^{'-}(dsb)$. (Even though the quark composition is identical, $\Xi_b^-$ and $\Xi_b^{'-}$ are distinct.) Three such $\Xi_b$ isodoublets are expected to exist, none of which are orbitally or radially excited, and which can be classified by the spin $j$ of the $us$ or $ds$ light di-quark and the spin-parity $J^p$ of the baryon: one with $j = 0$ and $J^p=\frac{1}{2}^{+}$ ($\Xi_b$), one with $j = 1$ and $J^p=\frac{1}{2}^{+}$ ($\Xi_b^{'}$), one with $j = 1$ and $J^p=\frac{3}{2}^{+}$ ($\Xi_b^*$), follows the same pattern as the well-known $\Xi_c$ states \cite{Olive2014}, and we therefore refer to these three iso-doublets as the $\Xi_b$, the $\Xi_{b}^{'}$ and $\Xi_{b}^{*}$. The spin-antisymmetric $J^p=\frac{1}{2}^+$ state, observed by multiple experiments\cite{PDG}  others should decay predominantly strongly through a $P$-wave pion transition ($\Xi_{b}^{',*} \rightarrow \Xi_{b} \pi$) if their masses are above the kinematic threshold for such a decay; otherwise they should decay electromagnetically ($\Xi_{b}^{',*} \rightarrow \Xi_{b} \gamma$) \cite{AaijPRL114}.

The new excited state $\Xi_b^-(6227)$ is also detected by LHCb collaboration \cite{AaijPRL2018}. Three $\Xi_{b}$ excited states are also detected during proton-proton (pp) collision at the LHC at $\sqrt{13}=$TeV, which are, $\Xi_b^-(6100)-$, $\Xi_b^-(6327)^0$ and $\Xi_b^-(6333)^0$ \cite{Sirunyan2021,Aaij2021}.

The masses of radial states of $\Xi_b$ and $\Xi_{b}^{'}$ baryons are listed in Table \ref{ta2} and \ref{ta3}, and the masses of orbital states of $\Xi_b$ and $\Xi_{b}^{'}$ baryons are listed in Table \ref{ta4} and \ref{ta5}. 
 $2S$ state is lesser by 55-60 MeV from Ref.\cite{Ebert2011} and it will decrease with increment of principal quantum number. For $1P$ state, there is a few MeV difference with Ref.\cite{Ebert2011}, but for higher excited orbital states, the difference is increasing as 120-140 MeV for $1D$ state and around 300 MeV for $1F$ state. The new state $\Xi_{b}(6227)^{0,-}$, $\Xi_{b}(6100)^{-}$, $\Xi_{b}(6327)^{0}$ and $\Xi_{b}(6333)^{0}$  is compatible for $2S_{\frac{1}{2}^{+}}$, $1P_{\frac{3}{2}^{-}}$, $1D_{\frac{3}{2}^{+}}$ and $1D_{\frac{5}{2}^{+}}$ respectively, as per our calculation and it is shown with bold numbers in Table \ref{ta2} and \ref{ta3}. By CMS collaboration, $\Xi_{b}(6100)^{-}$ state is predicted as $1P(\frac{3}{2}^{-})$\cite{Sirunyan2021} and the narrow doublet $\Xi_{b}(6327)^{0}$ and  $\Xi_{b}(6333)^{0}$ are predicted as $1D(\frac{3}{2}^{+})$ and $1D(\frac{5}{2}^{+})$ by LHCb collaboration\cite{Aaij2021}. Azizi\cite{Azizi2021} studied the resonance $\Xi_{b}(6227)^{-}$ and conclude that the resonace may be $2S$ or $1P$ excited state of $\Xi_{b}$. And He, Liang et. al.\cite{HeLiang2021} concluded that $\Xi_{b}(6227)^{-}$ may be $1P (\frac{3}{2}^{-}$) of $\Xi_b^{'}$.  As $\Xi_{b}(6100)$ is defined $1P$ state\cite{Sirunyan2021}, $\Xi_{b}(6227)^{-}$ state must be $2S$ state. For $\Xi_{b}^{'-}$ baryon, $1S$ and $2S$ states are quite near to the Refs. \cite{HeLiang2021} and \cite{Wei2017}. Our masses $1P$ state is in good agreement with Ref. \cite{Wei2017} and 93 MeV differ from Ref. \cite{{WangCPC2017}}. $1D$ and $1F$ states are in decrements of around 100 MeV and 250 MeV from Refs. \cite{HeLiang2021} and \cite{Wei2017} due to screening effect.
The suppression of our calculated mass from the other theoretical predictions is showing the screening effect. The $\Xi_b^0$ and $\Xi_b^-$ are very narrow states, they are considered as single state by taking the average of constituent quark masses of $u$ and $d$ quarks.

\begin{table}[h]
	\tbl{Predicted masses of radial states of $\Omega_{b}$ baryon (in GeV).}
	{\begin{tabular}{@{}cccccccccccccccc@{}} \toprule
			state & \multicolumn{2}{c}{Present} & PDG\cite{PDG}& \multicolumn{2}{c}{\cite{Ebert2011}} & \multicolumn{2}{c}{\cite{Yamaguchi2015}} & \cite{Jia2021} & \cite{Moosavi2020} & \cite{Wei2017} \\
			
			& $S=\frac{1}{2}$ & $S=\frac{3}{2}$ && $S=\frac{1}{2}$ & $S=\frac{3}{2}$ & $S=\frac{1}{2}$ & $S=\frac{3}{2}$  \\ \colrule
			$1S$ & 6.046 & 6.082 & 6.046 & 6.064 & 6.088 &6.081 & 6.102 & 6.051 & 6.098 & 6.048\\
			$2S$ & 6.438 & 6.462 && 6.450 & 6.461 & 6.472 & 6.478  & 6.489\\ 
			$3S$ & 6.740 & 6.753 && 6.804 & 6.811 &6.593 & 6.593 \\ 
			$4S$ & 7.022 & 7.030 && 7.091 & 7.096 \\ 
			$5S$ & 7.290 & 7.296 && 7.338 & 7.343 \\ 
			$6S$ & 7.546 & 7.549 & \\ 
			\botrule
		\end{tabular}\label{ta6}}	
\end{table}	

\subsection{$\Omega_b^-$ State}

\begin{table}
	\tbl{Predicted masses of orbital excited states of $\Omega_{b}$ baryon (in GeV).}
	{\begin{tabular}{@{}cccccccccccc@{}} \toprule
		state & $2S+1$ & $J^p$ & Present& PDG\cite{PDG} & \cite{Ebert2011} & \cite{Mutuk2020} & \cite{Jia2021} & \cite{Wei2017} & \cite{WangCPC2017} & \cite{Yoshida}&\cite{Juhi}\\  \colrule
	1P & 2 & $\frac{1}{2}^{-}$ & \textbf{6.344} &\textbf{6.330}& 6.339 & 6.314 & 6.342 && 6.248\\
	&   & $\frac{3}{2}^{-}$ & \textbf{6.341}& \textbf{6.316} & 6.340 & 6.330 && 6.325 & 6.207&&6.348\\
	& 4 & $\frac{1}{2}^{-}$ & 6.345 && 6.330 & 6.339 &&& 6.269 & 6.333\\
	&   & $\frac{3}{2}^{-}$ & \textbf{6.343}&\textbf{6.350 }& 6.331 & 6.342&&&& 6.336 \\
	&   & $\frac{5}{2}^{-}$ & \textbf{6.339}& \textbf{6.340} & 6.334 & 6.352&&&&6.345&6.362\\
	\hline   
	2P & 2 & $\frac{1}{2}^{-}$ & 6.596& & 6.710 && 6.724\\
	&   & $\frac{3}{2}^{-}$ & 6.594 & &6.705 &\\
	& 4 & $\frac{1}{2}^{-}$ & 6.597 & &6.706 &&&&& 6.740\\
	&   & $\frac{3}{2}^{-}$ & 6.595 & &6.699&&&&& 6.744\\
	&   & $\frac{5}{2}^{-}$ & 6.592 & &6.700 &&&&&6.728\\
	\hline   
	3P & 2 & $\frac{1}{2}^{-}$ & 6.829& & 7.009\\
	&   & $\frac{3}{2}^{-}$ & 6.827 & & 7.003\\
	& 4 & $\frac{1}{2}^{-}$ & 6.830 & &7.002 &&&&& 6.937\\
	&   & $\frac{3}{2}^{-}$ & 6.828 & &6.998 &&&&& 6.937\\
	&   & $\frac{5}{2}^{-}$ & 6.826 & &6.996 &&&&&6.919\\
	\hline      
	4P & 2 & $\frac{1}{2}^{-}$ & 7.044& & 7.265\\
	&   & $\frac{3}{2}^{-}$ & 7.043 && 7.258\\
	& 4 & $\frac{1}{2}^{-}$ & 7.043 && 7.257\\
	&   & $\frac{3}{2}^{-}$ & 7.043 && 7.250\\
	&   & $\frac{5}{2}^{-}$ & 7.042 && 7.251\\
	\hline

	1D & 2 & $\frac{3}{2}^{+}$ & 6.480 && 6.549&& 6.593&&&& 6.629\\
&   & $\frac{5}{2}^{+}$ & 6.476 && 6.529 &&& 6.590 && 6.561\\
& 4 & $\frac{1}{2}^{+}$ & 6.485 && 6.540\\
&   & $\frac{3}{2}^{+}$ & 6.482 && 6.530\\
&   & $\frac{5}{2}^{+}$ & 6.478 && 6.520&&&&&&6.638\\   
&   & $\frac{7}{2}^{+}$ & 6.472 && 6.517\\
\hline   
	2D & 2 & $\frac{3}{2}^{+}$ & 6.726 && 6.863 && 6.936\\
	&   & $\frac{5}{2}^{+}$ & 6.723 && 6.846 &&&&& 6.866\\
	& 4 & $\frac{1}{2}^{+}$ & 6.730 && 6.857\\
	&   & $\frac{3}{2}^{+}$ & 6.727 && 6.846\\
	&   & $\frac{5}{2}^{+}$ & 6.724 && 6.837\\   
	&   & $\frac{7}{2}^{+}$ & 6.720 && 6.834\\   
	\hline   

	3D & 2 & $\frac{3}{2}^{+}$ & 6.953 & \\
	&   & $\frac{5}{2}^{+}$ & 6.951  & \\
	& 4 & $\frac{1}{2}^{+}$ & 6.956  & \\
	&   & $\frac{3}{2}^{+}$ & 6.954  & \\
	&   & $\frac{5}{2}^{+}$ & 6.951  & \\   
	&   & $\frac{7}{2}^{+}$ & 6.948  & \\
	\hline   
	4D & 2 & $\frac{3}{2}^{+}$ & 7.164 & \\
	&   & $\frac{5}{2}^{+}$ & 7.162 & \\
	& 4 & $\frac{1}{2}^{+}$ & 7.166 & \\
	&   & $\frac{3}{2}^{+}$ & 7.164 & \\
	&   & $\frac{5}{2}^{+}$ & 7.162 & \\   
	&   & $\frac{7}{2}^{+}$ & 7.160 & \\   
\hline
\end{tabular}\label{ta7}}
{\it Table 9.}	$(${\it Continued}$)$
\end{table}

\begin{table}
\centering
{\begin{tabular}{@{}ccccccccccccc@{}} \toprule
state & $2S+1$ & $J^p$ & Present & \cite{Ebert2011} & \cite{Mutuk2020} & \cite{Jia2021} & \cite{Wei2017} & \cite{WangCPC2017} & \cite{Yoshida}&\cite{Juhi}\\  \colrule
	1F & 2 & $\frac{5}{2}^{-}$ & 6.602& & 6.771&& 6.817\\
	&   & $\frac{7}{2}^{-}$ & 6.596 && 6.736 &&&6.844\\
	& 4 & $\frac{3}{2}^{-}$ & 6.609 && 6.763\\
	&   & $\frac{5}{2}^{-}$ & 6.604 && 6.737\\
	&   & $\frac{7}{2}^{-}$ & 6.597 && 6.719&&&&&6.899\\   
	&   & $\frac{9}{2}^{-}$ & 6.590 && 6.713&&&&&6.903\\   
	\hline   
	2F & 2 & $\frac{5}{2}^{-}$ & 6.843 && && 7.132\\
	&   & $\frac{7}{2}^{-}$ & 6.839 & \\
	& 4 & $\frac{3}{2}^{-}$ & 6.848 & \\
	&   & $\frac{5}{2}^{-}$ & 6.845 & \\
	&   & $\frac{7}{2}^{-}$ & 6.840 & \\   
	&   & $\frac{9}{2}^{-}$ & 6.835 &&&&& \\      
	\noalign{\smallskip}
			
			\botrule
		\end{tabular}}	
\end{table}	

$\Omega_b$ is the heaviest member of the singly bottom family, consist of one bottom (heavy) quark and two strange (light) quarks with strangeness of -2. The measurements of the mass and lifetime of the $\Omega_{b}^-$ baryon have been done using the various decay modes.  The $\Omega_{b}^-$ baryon detected by LHCb collaboration \cite{AaijPRD2016},  CDF collaboration \cite{CDF2009} and $D0$ collaboration \cite{D02008} (Fermi lab).
The radial and orbital masses of $\Omega_b$ baryon is listed in Table \ref{ta8} and \ref{ta9} respectively. The mass of $1S$ state of $\Omega_{b}^{-}$ is quite near to the compared Refs. \cite{Ebert2011,Jia2021,Moosavi2020,Wei2017} with difference of 5-50 MeV. The masses of $2S-6S$ states are also near to the Refs. \cite{Ebert2011,Jia2021,Moosavi2020,Wei2017}.\\

\noindent  For $1P$ states, our calculated masses are in accordance with the newly detected four narrow states. We assigned the $J^p$ values to the states, which are shown with bold in Table 7.  One of the four newly observed narrow states, $\Omega_b(6316)$ can be assigned as $1^2P_{\frac{3}{2}}$, which is 25 MeV lesser than our calculated mass for the particular state. The other three states, $\Omega_b(6330)$, $\Omega_b(6340)$ and $\Omega_b(6350)$ can be assigned as further hyperfine states of $1P$, and they are 11 MeV, 5 MeV and 3 MeV differ from our calculated masses respectively. These narrow states are also defined by Z. G. Wang\cite{ZGWang} as: $\Omega_b(6316) \rightarrow 1P(\frac{3}{2}^-)$, $\Omega_b(6330) \rightarrow  1P(\frac{1}{2}^-)$, $\Omega_b(6340) \rightarrow  1P(\frac{5}{2}^-)$ and $\Omega_b(6350) \rightarrow  1P(\frac{3}{2}^-)$, using full QCD sumrules. Our results are also matching with Refs. \cite{Ebert2011,Mutuk2020} and as the principal quantum number increases, our results are suppressed around 200 MeV from the compared theoretical predictions due to screening effect. The similar effect can be observed for $D$ and $F$ state also.

\subsection{Regge Trajectories}

\begin{figure}
	\centerline{\includegraphics[width=8.0cm]{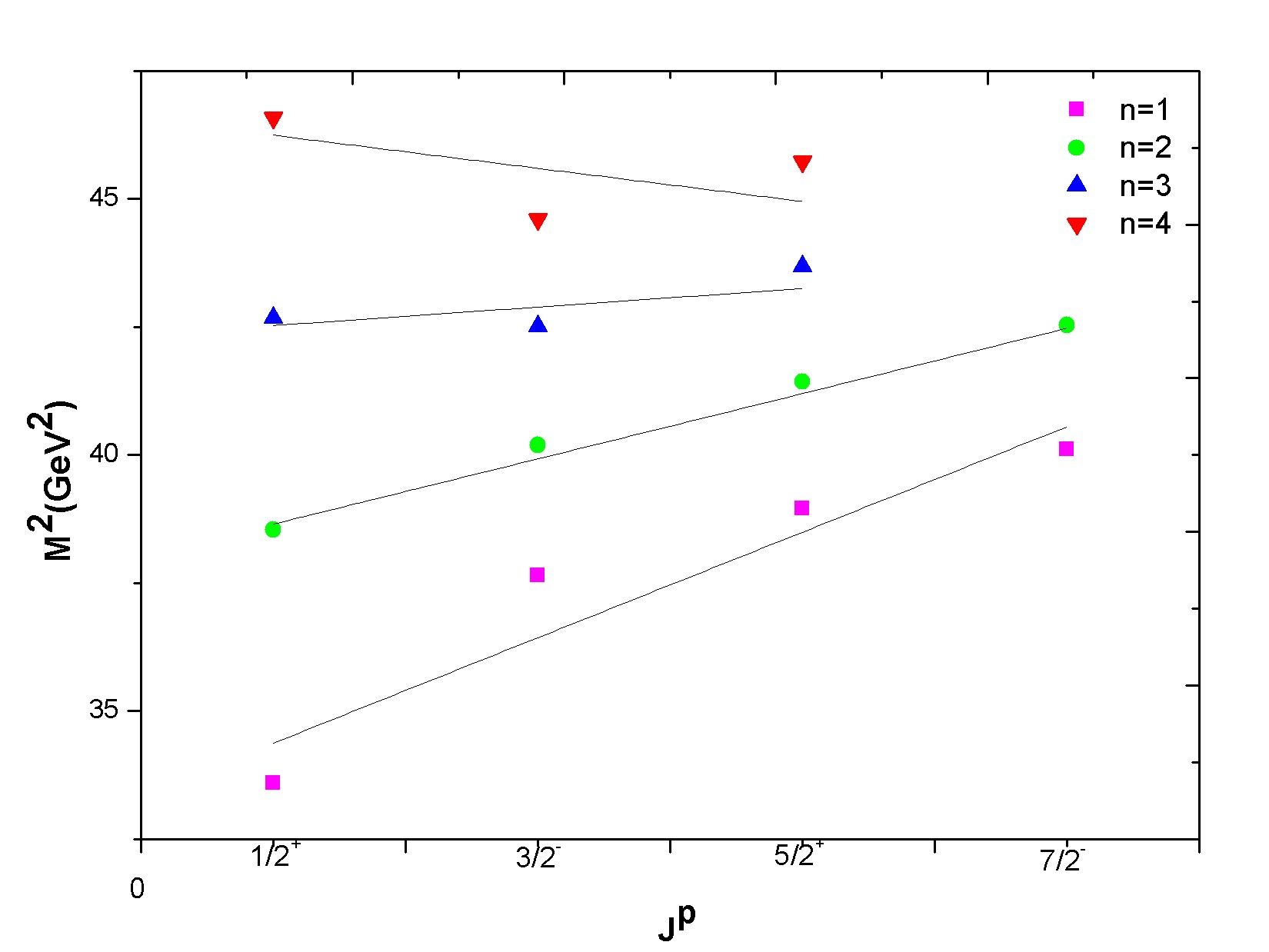}}
	\caption{Regge trajectory of $\Xi_{b}$ baryon in $(J, M^2)$ plane \label{f1}}
\end{figure}

\begin{figure}
	\centerline{\includegraphics[width=8.0cm]{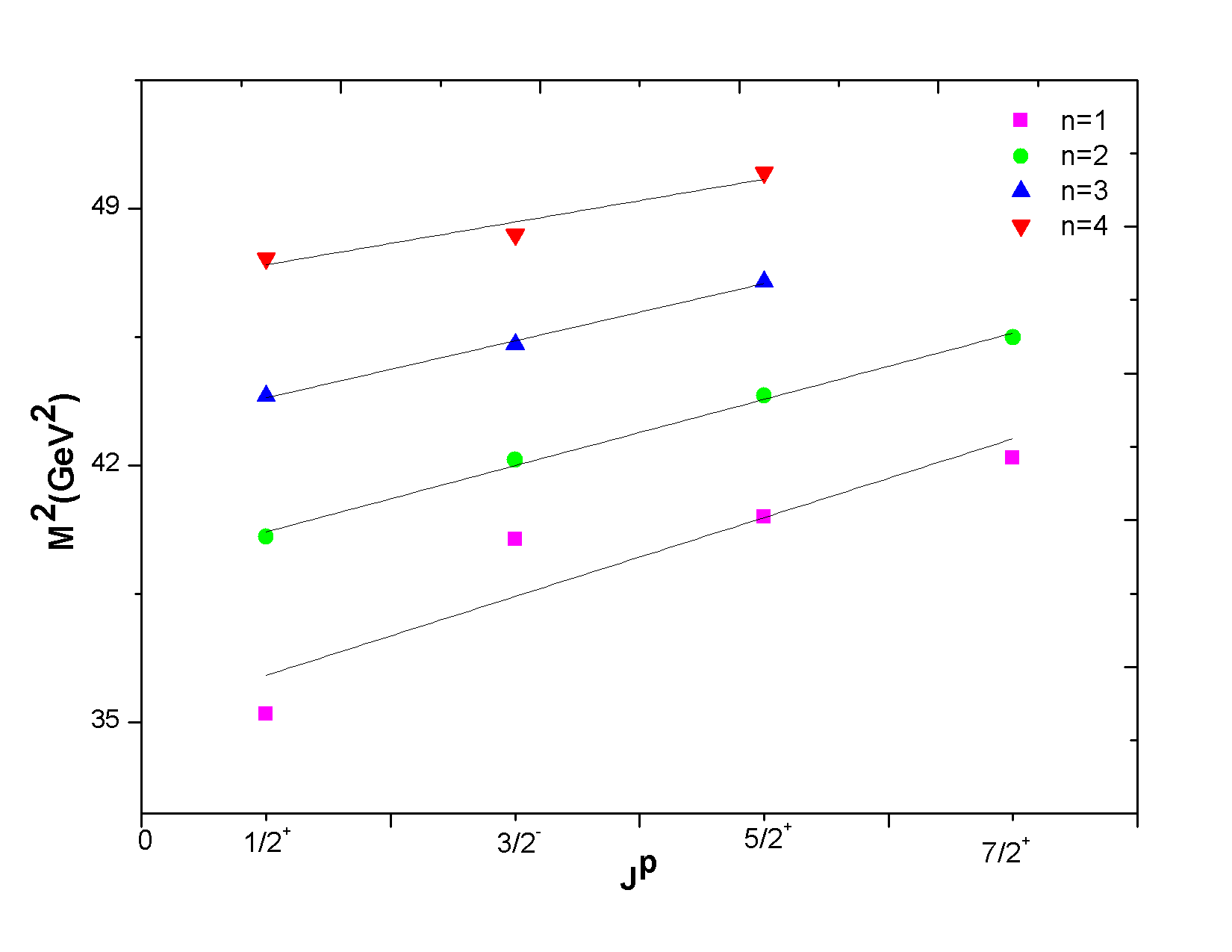}}
	\caption{Regge trajectory of $\Xi_{b}^{'}$ baryon in $(J, M^2)$ plane \label{f2}}
\end{figure}

\begin{figure}
	\centerline{\includegraphics[width=8.0cm]{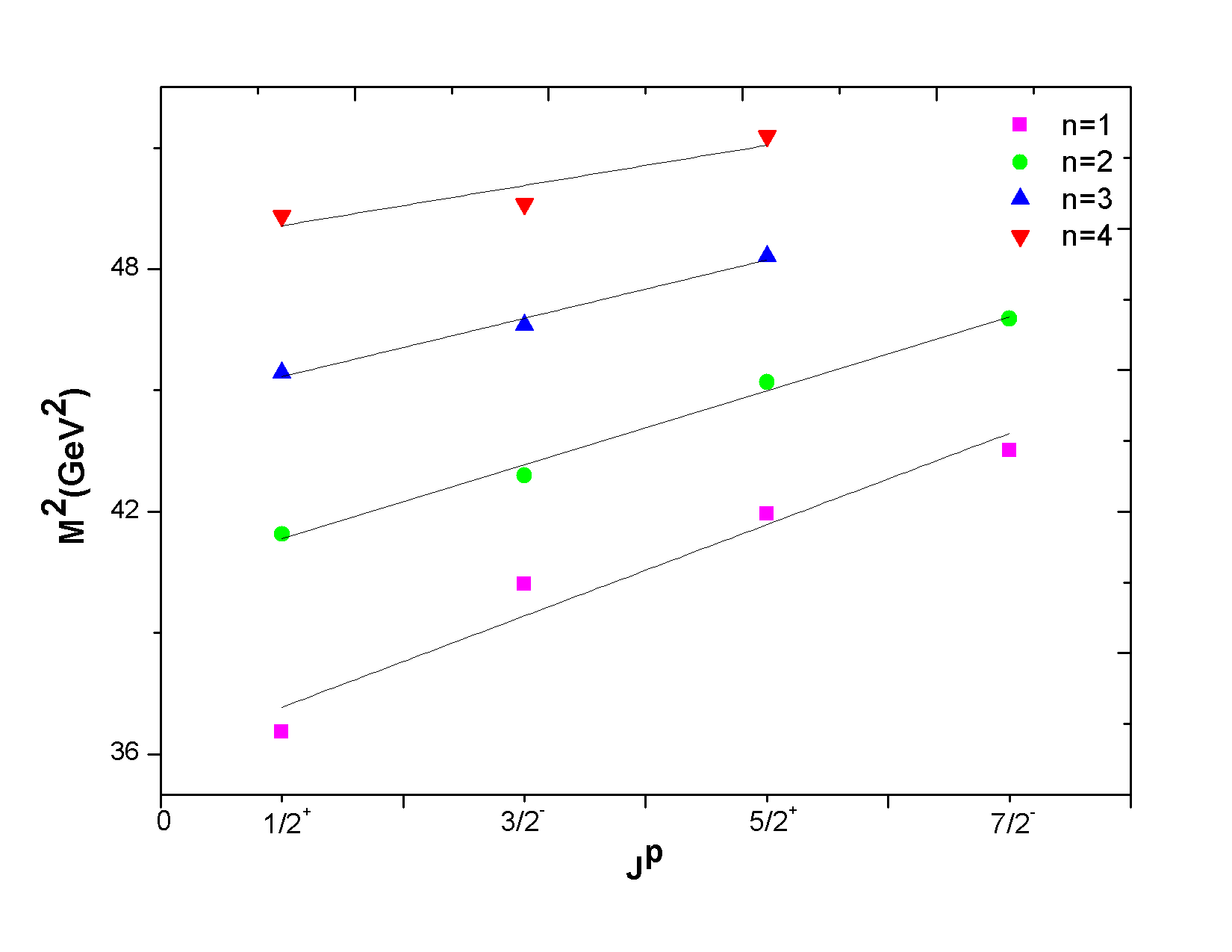}}
	\caption{Regge trajectory of $\Omega_{b}$ baryon in $(J, M^2)$ plane \label{f3}}
\end{figure}
The Regge trajectories are plotted for $\Xi_b$, $\Xi_b$' and $\Omega_b$ baryons. The trajectories are drawn in the ($J, M^2$) plane using calculated masses for baryonic systems. The linearity of the Regge trajectories is a useful probe for predicting the masses of higher excited states, as well as for defining quantum number if the resonance mass is known. The Regge trajectories have been plotted for the states $J^p = \frac{1}{2}^+, \frac{3}{2}^-, \frac{5}{2}^+ $ and $ \frac{7}{2}^-$ using the equation\cite{Ebert2011},
\begin{equation}
	J=\alpha M^2 + \alpha_0
\end{equation}

The Regge trajectories of $\Xi_b$, $\Xi_b^{'}$ and $\Omega_b$ are shown in Figure \ref{f1}, \ref{f2} and \ref{f3} respectively. Regge trajectories of $\Xi_b$ baryons are linear but the Regge lines intersect each other for higher states which shows the effect of screened potential.

\section{Properties}
\label{section4}
\subsection{Magnetic Moments}
The magnetic moment of a baryon is an intrinsic property caused by the spin of its constituents. The effective mass of the ground state, individual charge, and spin-flavor arrangement of the constituent quarks are used to calculate the magnetic moment of the ground state of the singly heavy bottom-strange baryon. The expression of magnetic moment of baryon can be obtained by operating the expectation value equation \cite{BPatel},

\begin{equation}
	\mu_B=\sum_q \langle \Phi_{sf}|\hat{\mu}_{qz}|\Phi_{sf} \rangle;  \quad \quad q=u, d, b
\end{equation}
where, $\Phi_{sf}$ is the spin-flavor wave-function of the baryon and $\hat{\mu}_{qz}$ is the magnetic moment operator. The magnetic moment of individual quark is given by \cite{Majethiya2009},
\begin{equation}
	\mu_q=\frac{e_q}{2m_q^{eff}}\cdot \sigma_q
\end{equation}
\noindent where, $e_q$ and $\sigma_q$ are charge and spin of the individual constituent quark of the baryonic system respectively and $m_q^{eff}$ is the effective mass of constituent quark. 
The magnetic moment of ground states of all baryonic systems are being calculated and shown in the Table \ref{ta8}. Our results are also compared with different approaches. The obtained values are differ only by 0.1 to 0.5 $\mu_N$ (nuclear magneton) from Ref.\cite{Majethiya2009}. 

\begin{table}[h]
	\tbl{Magnetic moment of ground state of singly heavy strange bottom baryons (in $\mu_N$).}
	{\begin{tabular}{@{}cccccccccccccc@{}} \toprule
		Baryon& $J^p$ & Expression  & Present & \cite{Majethiya2009} & \cite{Dhir} & \cite{Ghalenovia2014} & \cite{BPatel} & \cite{Bernotas} & \cite{Franklin}\\ \colrule
	$\Xi_{b}^0$ & $\frac{1}{2}^+$ & $\frac{2}{3}\mu_u+\frac{2}{3}\mu_s-\frac{1}{3}\mu_b$ & 0.798 & 0.799 & -0.062 & - & - & -0.100 & -0.110 \\
	$\Xi_{b}^-$ & $\frac{1}{2}^+$ & $\frac{2}{3}\mu_d+\frac{2}{3}\mu_s-\frac{1}{3}\mu_b$  & -0.963 & -0.958 & -0.062 & - & - & -0.063 & -0.050\\
	$\Xi_{b}^{'-}$ & $\frac{1}{2}^+$ & $\frac{2}{3}\mu_d+\frac{2}{3}\mu_s-\frac{1}{3}\mu_b$ & -0.941 && -0.913 & - & -\\ 
	$\Omega_{b}^-$ & $\frac{1}{2}^+$ & $\frac{4}{3}\mu_s-\frac{1}{3}\mu_b$ & -0.761 & -0.752 & -0.741 & - & - & -0.545 & -0.790\\
	$\Xi_{b}^{*0}$ & $\frac{3}{2}^+$ & $\mu_u+ \mu_s+ \mu_b$ & 1.101 & 1.083 & 1.031 & 1.136 & 1.041 & 0.690 & 1.190\\
	$\Xi_{b}^{*-}$ & $\frac{3}{2}^+$ & $\mu_d+ \mu_s+ \mu_b$ & -1.499 & -1.505 & -1.454 & -1.621 & -1.095 & -1.088 & -1.600\\
	$\Xi_{b}^{*'-}$ & $\frac{3}{2}^+$ & $\mu_d+ \mu_s+ \mu_b$ & -1.498 \\
	$\Omega_{b}^{*-}$ & $\frac{3}{2}^+$ & $2\mu_s+ \mu_b$ & -1.236 & -1.222 & -1.201 & -1.380 & -1.199 & -0.919 & -1.280\\		
			\botrule
		\end{tabular}\label{ta8}}	
\end{table}	

\subsection{Radiative Decays}
The higher energy baryonic state decays into lower energy state, there is an emission of $\gamma$ radiation. The decay width for this radiative transition is \cite{ShahCPC2016} ,
\begin{equation}
	\Gamma=\frac{k^3}{4\pi}\frac{2}{2J+1}\frac{e^2}{2m_p^2}\mu_{B\rightarrow B'}^2
\end{equation}
	where, $k$ is photon energy, $J$ is total angular momentum of the initial baryonic state, $m_p$ is the mass of proton (in MeV) and $\mu_{B\rightarrow B'}$ is transition magnetic moment for the particular radiative decay ($B$ is initial baryonic state and $B'$ is final baryonic state).

\begin{table}[h]
	\tbl{Radiavtive Transition magnetic moment}
	{\begin{tabular}{@{}cccccccccc@{}} \toprule
	Transition & Expression & Present &  ecqm\cite{Majethiya2009}  & cqm \cite{Majethiya2009}  \\  \colrule
	
	$\Xi_{b}^{*0}\rightarrow \Xi_{b}^0 \gamma$ & $\frac{\sqrt{2}}{\sqrt{3}}(\mu_u-\mu_s)$ & 1.896 &1.036 &1.222  \\
	$\Xi_{b}^{*-}\rightarrow \Xi_{b}^- \gamma$ & $\frac{\sqrt{2}}{\sqrt{3}}(\mu_d-\mu_s)$ & -0.230 & -0.124 & -0.147 \\
	$\Xi_{b}^{*'-}\rightarrow \Xi_{b}^{'-} \gamma$ & $\frac{\sqrt{2}}{\sqrt{3}}(\mu_d-\mu_s)$ & -0.227 \\
	$\Omega_{b}^{*-}\rightarrow \Omega_{b}^- \gamma$ & $\frac{\sqrt{2}}{3}(\mu_s-\mu_b)$ & -0.246 & -0.523 & -0.632 \\		
			\botrule
		\end{tabular}\label{ta9}}	
\end{table}	

\begin{table}[h]
	\tbl{Radiavtive decay width}
	{\begin{tabular}{@{}cccccccccc@{}} \toprule
			Transition & Expression &Present & ecqm\cite{Majethiya2009} &cqm\cite{Majethiya2009} \\ \colrule
			$\Xi_{b}^{*0}\rightarrow \Xi_{b}^0 \gamma$ & $\frac{\sqrt{2}}{\sqrt{3}}(\mu_u-\mu_s)$ & 71.04 & 18.79 & 26.14 \\
			$\Xi_{b}^{*-}\rightarrow \Xi_{b}^- \gamma$ & $\frac{\sqrt{2}}{\sqrt{3}}(\mu_d-\mu_s)$ &  0.99 & 0.09 & 0.13\\
			$\Xi_{b}^{*'-}\rightarrow \Xi_{b}^{'-} \gamma$ & $\frac{\sqrt{2}}{\sqrt{3}}(\mu_d-\mu_s)$ &  0.22\\
			$\Omega_{b}^{*-}\rightarrow \Omega_{b}^- \gamma$ & $\frac{\sqrt{2}}{3}(\mu_s-\mu_b)$ & 0.01 & 0.03 & 0.04\\		
			\botrule
		\end{tabular}\label{ta10}}	
\end{table}	

The transition magnetic moment is expressed as, 
	\begin{equation}
		\mu_{B\rightarrow B'}=\langle \Phi_B|\hat{\mu_{B\rightarrow B'}}|\Phi_{B'}\rangle    
	\end{equation}
Here, $\Phi_B$ and $\Phi_B'$ are the wave functions of initial and final baryonic states respectively.
The transition magnetic moment and radiative decay widths are listed in Table \ref{ta9} and \ref{ta10}. Our results for transition magnetic moment and transition decay width are in range of the results shown in Ref. \cite{Majethiya2009}. As the $\Xi_{b}^{'-}$ baryon is not much explored, there is no data to compare for its properties like magnetic moment, transition magnetic moment and radiative decay width.

\subsection{Strong Decays}

The singly strange bottom baryons contains one bottom quark(b), one strange quark(s) and one of the two light quarks(u and d). This structure gives an important ground 
for testing the heavy quark symmetry of the bottom quark(heavy quark) and the chiral symmetry of the light quarks, due to this fact strong decays of these baryons are studied in the framework of the heavy hadron chiral perturbation theory (HHChPT)\cite{Pirjol,ChengPRD2007,ChengPRD2017,ChengPRD2015}.

Strong decay transitions are categorized in three ways according to its initial and final states: $P$-wave transition, $S$-wave transition and $D$-wave transition. The decay widths for  $P$-wave, $S$-wave and $D$-wave transitions have been calculated in this work using the Lagrangian given in Ref.\cite{Pirjol}; where, $P$-wave transition is transition between $s$-wave baryons, $S$-wave transition is transition between $p$-wave ($J^P=\frac{1}{2}$) to $s$-wave baryons and $D$-wave transition is  transition between $p$-wave ($J^P=\frac{3}{2}$,$\frac{5}{2}$) to $s$-wave baryons.

The expression of the decay width for $P$-wave transitions $\Xi_{b}^{0}(1^2S_{\frac{3}{2}}) \rightarrow \Xi_{b}^{0} \pi$, $\Xi_{b}^{-}(1^2S_{\frac{3}{2}}) \rightarrow \Xi_{b}^{-} \pi$ and $\Xi_{b}^{'-}(1^2S_{\frac{3}{2}}) \rightarrow \Xi_{b}^{-} \pi$ respectively is \cite{ChengPRD2007},
	\begin{equation}
	\Gamma=\frac{a_1^2}{8\pi f_{\pi}^2} \frac{M_{\Xi_{b}^{0,-}}}{M_{\Xi_{b}^{0,-,'-}}(1^2S_{\frac{3}{2}})} p_{\pi}^3
\end{equation}

And the expressions of the decay width for $P$-wave transitions $\Xi_{b}^{0}(1^2S_{\frac{3}{2}}) \rightarrow \Xi_{b}^{-} \pi$, $\Xi_{b}^{-}(1^2S_{\frac{3}{2}}) \rightarrow \Xi_{b}^{0} \pi$ and $\Xi_{b}^{'-}(1^2S_{\frac{3}{2}}) \rightarrow \Xi_{b}^{0} \pi$ respectively is \cite{ChengPRD2007},
\begin{equation}
	\Gamma=\frac{a_1^2}{4\pi f_{\pi}^2} \frac{M_{\Xi_{b}^{-,0}}}{M_{\Xi_{b}^{0,-,'-}}(1^2S_{\frac{3}{2}})} p_{\pi}^3
\end{equation}

where, strong coupling constant $a_1=0.565$ \cite{ChengPRD2007}, the pion decay constant $f_{\pi}=132$ MeV. Masses of initial and final baryon state have been taken as shown in Section \ref{section3}. $p_{\pi}^3$ represents the $P$-wave transition momentum \cite{ChengPRD2017} (for two body decay $x \rightarrow y + \pi \slash K$)  which can be expressed as,
	\begin{equation}
	p_{\pi \slash K} =\frac{1}{2m_x}\sqrt{[m_{x}^2-(m_y+m_\pi \slash m_K)^2][m_{x}^2-(m_y-m_\pi \slash m_K)^2]}   
\end{equation}

	The expression of decay width for $S$-wave transition $\Xi_{b}^{0}(1^2P_{\frac{1}{2}}) \rightarrow \Xi_{b}^{'-} \pi$, 

\begin{equation}
	\Gamma=\frac{b_1^2}{4\pi f_{\pi}^2} \frac{M_{\Xi_{b}^{'-}}}{M_{\Xi_{b}^{0}(1^2P_{\frac{1}{2}})}} E_{\pi}^2 p_{\pi}
\end{equation}

The decay width for $S$-wave transition $\Omega_{b}^{-}(1^2P_{\frac{1}{2}}) \rightarrow \Xi_{b}^{0} K^-$ is, 
\begin{equation}
	\Gamma=\frac{b_2^2}{4\pi f_{\pi}^2} \frac{M_{\Xi_{b}^{0}}}{M_{\Omega_{b}^{-}(1^2P_{\frac{1}{2}})}} E_{K}^2 p_{K}
\end{equation}
For this transition, $b_2=\sqrt{3}b_1$ \cite{ChengPRD2017} and $b_1=0.553$ \cite{ChengPRD2017} for this particular transition.

The expression of decay width for $D$-wave transition $\Xi_{b}^{0}(1^2P_{\frac{3}{2}}) \rightarrow \Xi_{b}^{*0} \pi$ is, 

\begin{equation}
	\Gamma=\frac{b_1^2}{8\pi f_{\pi}^2} \frac{M_{\Xi_{b}^{*0}}}{M_{\Xi_{b}^{0}(1^2P_{\frac{3}{2}})}} E_{\pi}^2 p_{\pi}
\end{equation}

And for $D$-wave transition $\Xi_{b}^{0}(1^2P_{\frac{3}{2}}) \rightarrow \Xi_{b}^{*-} \pi$, 

\begin{equation}
	\Gamma=\frac{b_1^2}{4\pi f_{\pi}^2} \frac{M_{\Xi_{b}^{*0}}}{M_{\Xi_{b}^{-}(1^2P_{\frac{3}{2}})}} E_{\pi}^2 p_{\pi}
\end{equation}
Here, the coupling constant $b_1=0.63$\cite{ChengPRD2015}, $p_{\pi}$ presents $S$-wave transition momentum and $E_{\pi} \approx m_{\pi}$ for the single pion at rest.

\begin{table}
	\tbl{Strong decay widths of singly heavy bottom-strange baryons.}
	{\begin{tabular}{@{}cccccccccc@{}} \toprule
			Decay channel & decay width (in MeV) & \cite{MajethiyaThesis}& \cite{Limphirat} & \cite{Mao2015}\\  \colrule
			$P$-wave transitions\\
			\hline
			$\Xi_b^{*0}\rightarrow \Xi_b^{0}+ \pi^0$ & 0.339 & 0.20 & & 1.3 \\
			$\Xi_b^{*0}\rightarrow \Xi_b^{-}+ \pi^
		+$ & 0.398 & & $2.4^{+0.1}_{-0.2}$ \\
			$\Xi_b^{*-}\rightarrow \Xi_b^{-}+ \pi^0$ & 0.265 & 0.40 & & 1.3 \\
			$\Xi_b^{*-}\rightarrow \Xi_b^{0}+ \pi^-$ & 0.839 & & $2.4^{+0.1}_{-0.2}$ \\
			$\Xi_b^{'*-}\rightarrow \Xi_b^{-}+ \pi^0$ & 0.365\\
			$\Xi_b^{'*-}\rightarrow \Xi_b^{0}+ \pi^-$ & 1.077\\
			\hline
			$S$-wave Transitions\\
			\hline
			$\Xi_b^{0}(1P)\rightarrow \Xi_b^{'-}+ \pi^+$ & 4.469 \\
			$\Xi_b^{0}(1P)\rightarrow \Xi_b^{*0}+ \pi^0$ & 1.783 \\
			$\Xi_b^{0}(1P)\rightarrow \Xi_b^{*-}+ \pi^+$ & 3.399 \\
			$\Omega_b^{-}(1P)\rightarrow \Xi_b^{0}+ \bar{K}^-$ & 445.1 \\
			\hline
			$D$-wave transitions\\
			\hline
			$\Xi_b^{-}(1P)\rightarrow \Xi_b^{-}+ \pi^0$ & 1.517 \\
			$\Omega_b^{-}(1P)\rightarrow \Xi_b^{0}+ \bar{K}^-$ & 0.489 \\			
			\botrule
		\end{tabular}\label{ta11}}	
\end{table}

The expression of decay width for $D$-wave transitions $\Xi_{b}^{-}(1^2P_{\frac{3}{2}}) \rightarrow \Xi_{b}^{-} \pi$ and $\Omega_{b}^{-}(1^2P_{\frac{3}{2}}) \rightarrow \Xi_{b}^{0} K^-$is ,
\begin{equation}
	\Gamma=\frac{4b_3^2}{15\pi f_{\pi}^2} \frac{M_{\Xi_{b}^{-,0}}}{M_{\Xi_{b}^{-}(1^2P_{\frac{3}{2}}) \slash \Omega_{b}^{-}(1^2P_{\frac{3}{2}})}} p_{\pi \slash K}^5
\end{equation}
here, coupling constant $b_3$ \cite{ChengPRD2007,ChengPRD2017}=$0.4 \times 10^{-3} MeV^{-1}$  and $p_{\pi \slash K}^5$ presents $D$-wave transition momentum for pion or kaon.

The calculated decay widths for $P$-wave, $S$-wave and $D$-wave transitions of  singly bottom-strange baryons have been listed in Table \ref{ta11} comparing with other references.

\section{Summary}
The Hypercentral Constituent Quark Model(hCQM) was used to calculate the masses of radial and orbital states of singly heavy bottom-strange baryons. The color-Coulomb potential has been taken along with the confining screening potential. The calculated mass spectra are compared with various approaches and our results are in agreement with them with few MeV mass difference. In general, masses of radial (1S-3S) states and $1P$ state are close to other theoretical predictions. The screening effect can be seen at higher excited states due to the screened potential (confining potential). Above the $2P$ state, screening of 100 MeV or higher are observed in all baronic systems. Screening effect is also visible from the Regge trajectories (Figs. \ref{f1}-\ref{f3}) for higher spin states. \\
\begin{enumerate}

\item The ground state masses of spin states $\frac{1}{2}$ and $\frac{3}{2}$ are in good agreement with the experimental as well as the other theoretical predictions for all bottom strange baryons (See Table \ref{ta2}, \ref{ta3} and \ref{ta6}).

\item In terms of assigning $J^p$ values, the resonance state $\Xi_b(6227)^{0,-}$ corresponds to our $\Xi_b^{'}$ baryon 1P state with $(\frac{3}{2}^-)$. Also, $\Xi_b(6100)^{-}$ is assigned as $1P (\frac{3}{2}^-) $, $\Xi_b(6327)^{0}$ is assigned as $1D (\frac{3}{2}^+) $ and $\Xi_b(6333)^{0}$ is assigned as $1D (\frac{5}{2}^+) $ state with mass difference of 35 MeV, 84 MeV and 93 MeV respectively, wih experimental masses. The difference between our predicted data and the experimentally observed resonance in $1D$ doublet is around 90 MeV, which is due to screening effect in higher orbital quantum number states.

\item 
Two states $\Omega_b^-(6316)$ and  $\Omega_b^-(6330)$ are identified as (spin 1/2) $1P(\frac{3}{2}^-)$ and $1P(\frac{1}{2}^-)$  states with mass difference of 25 MeV and 14 MeV respectively. Other two states $\Omega_b^-(6350)$ and  $\Omega_b^-(6340)$ identified as (Spin 3/2) $1P(\frac{3}{2}^-)$ and $1P(\frac{5}{2}^-)$ states with the mass difference of 7 MeV and 1 MeV respectively. 

\item $\chi^2$/d.o.f or the goodness of fit test is a statistical test, which gives an idea how near our predicted mass values to experimental values or true values. As the value of $\chi^2$/d.o.f is large, our predicted masses contain large uncertainity and the theoretical model is not fit for the calculation and the small value of $\chi^2$/d.o.f indicates the small uncertainity. The goodness of fit test performed as per the expression: $\frac{\chi^2}{d.o.f.}=\Sigma \left(\frac{(observed-experimental)^2}{experimental}\right)$ \cite{dof}. In this work, the value of $\chi^2$/d.o.f is $2.743 \times 10^{-3}$ for $\Xi_{b}$, $4.015 \times 10^{-6}$ for $\Xi_{b}^{'}$ and $1.299 \times 10^{-4}$ for $\Omega_b$. From the given values of $\chi^2$/d.o.f for $\Xi_{b}$ and $\Omega_b$, it can be concluded that our data has less uncertainity and the model is perfactly fit for our calculation.

\begin{figure}
	\centerline{\includegraphics[width=14.0cm]{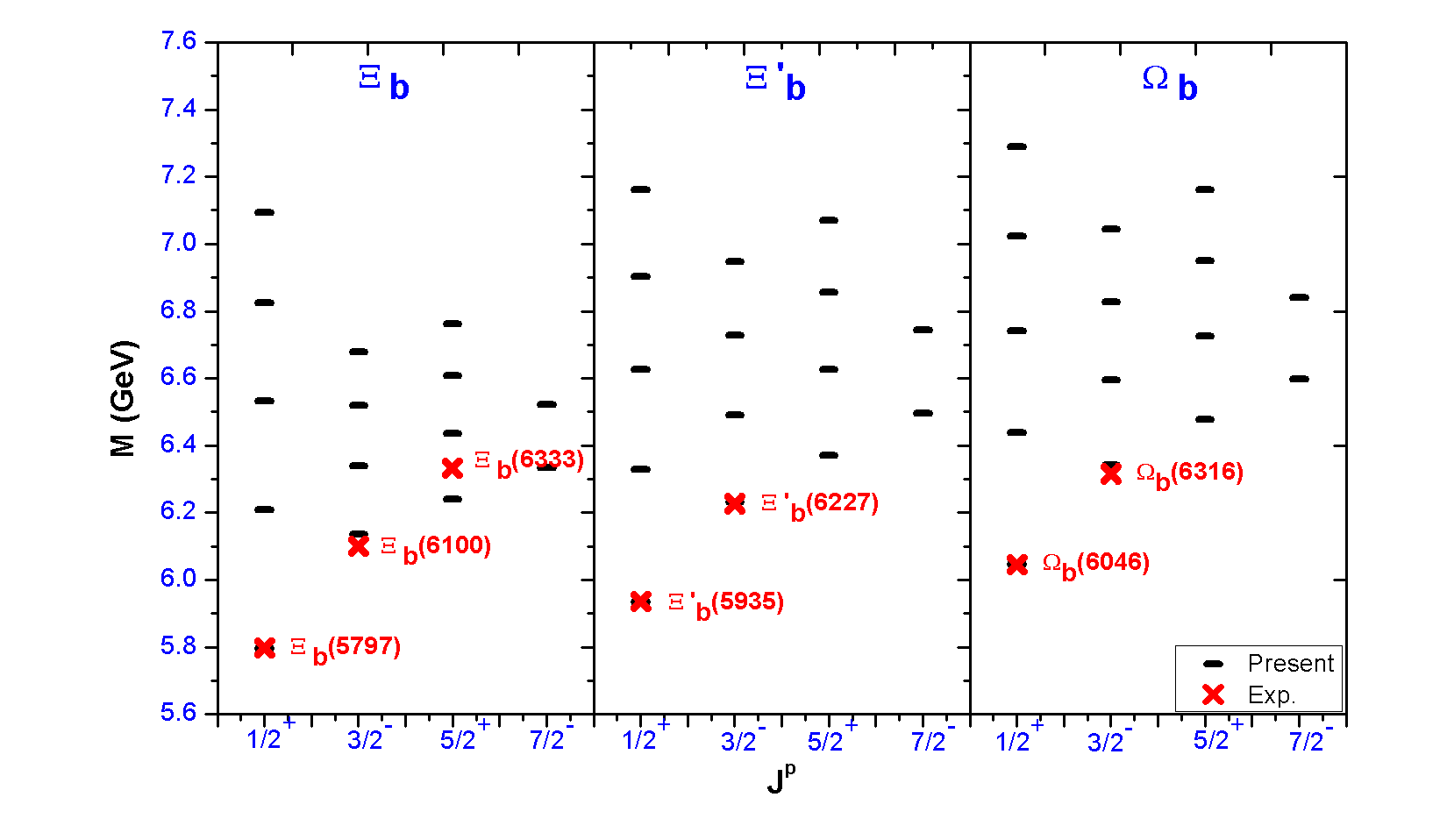}}
	\caption{Mass spectra representation of singly heavy bottom-strange baryons with experimentally detected states \label{f4}}
\end{figure}

\item The magnetic moments and transition magnetic moments of singly heavy bottom-strange baryons are calculated in constituent quark model. Magnetic moments of all baryons are reasonably close to  Ref. \cite{Majethiya2009}, but not same in case of transition magnetic moment. 
\item The strong decay width of various channels (P-wave, S-wave and D-wave of singly heavy bottom-strange baryons) are analyzed in the heavy hadron chiral perturbation theory(HHChPT), in absence of any experimental result we have compared our results to Ref.\cite{MajethiyaThesis} and Ref.\cite{Limphirat} (see Table \ref{ta11}). All results in table are not in mutually agreement with each other. More efforts are required for the decay properties of these baryons(as they are very important for the internal structure of baryons), in both the fronts, experimental as well as theoretical to study these properties to reach out for any conclusion.

\end{enumerate}
In this study, our aim is satisfied for the assignment of $J^P$ value of the experimentally measured resonance states, which can be seen in Figure \ref{f4} and model is successful in the determination of the various decay properties of singly heavy bottom-strange baryons. We would like to extend this model for doubly and triply heavy baryons in near future. \\

	

\begin{thebibliography}{00}  
		
		
	
	\bibitem{PDG} P. A. Zyla et al. (Particle Data Group), Prog. Theor. Exp. Phys. 2020, 083C01 (2020).
	
	\bibitem{AaijPRD2021} LHCb Collaboration (R. Aaij et. al.), {\it Phys. Rev. D} {\bf 103} 1, 012004 (2021).
	
	\bibitem{AaijPRL2018} LHCb Collaboration, (R. Aaij et. al.), {\it Phys. Rev. Lett.} {\bf 121}, 072002 (2018).
	
	\bibitem{AaijPRL2020} LHCb collaboration, (R. Aaij et. al.), {\it  Phys.Rev.Lett.} {\bf 124} 8, 082002 (2020).
	
	\bibitem{Aaij2021} LHCb collaboration, (R. Aaij et. al.), arXiv:2110.04497v1 [hep-ex] (2021).
	
	\bibitem{Sirunyan2021} CMS Collaboration, (A. M. Sirunyan et. al.), \textit{Phys. Rev. Lett.} \textbf{126}, 252003 (2021).
	
	\bibitem{Jia2019}D. Jia, W. N. Liu and A. Hosaka,\textit{ Phys. Rev. D} \textbf{101}, 034016 (2020).

\bibitem{Azizi2021} K. Azizi, Y. Sarac and H. Sundu,  \textit{J. High Energ. Phys.} \textbf{244} (2021). 
	
	\bibitem{Xiao2020} L. Y. Xiao, K.L. Wang, M. S. Liu, et al., \textit{Eur. Phys. J. C} \textbf{80}, 279 (2020).
	
	\bibitem{Jia2021} D. Jia, J. H. Pan, C. Q. Pang, {\it Eur. Phys. J. C} {\bf 81}, 434 (2021).
	  \bibitem{Moosavi2020}
   S. M. Moosavi Nejada, A. Armat, {\it Eur. Phys. J. A} {\bf 56}, 287 (2020).
	
		\bibitem{HeLiang2021}
   H. Z. He, W. Liang et. al., {\it Sci. China Phys. Mech. Astron.} {\bf 64}, 261012 (2021).

	
	\bibitem{Yamaguchi2015}Y. Yamaguchi, S. Ohkoda, et. al., {\it Phys. Rev. D} {\bf 91}, 034034 (2015).
	\bibitem{WangCPC2017}
   Z. Y. Wang, J. J. Qi et. al., {\it Chin. Phys. C} {\bf 41}, 093103 (2017). 
   
   \bibitem{Mutuk2020}
   H. Mutuk, {\it Eur. Phys. J. A} {\bf 56}, 146 (2020). 

\bibitem{Ebert2011}
	D. Ebert, R. N. Faustov and V. O. Galkin, {\it Phys. Rev. D} {\bf 84}, 014025 (2011).
	
     \bibitem{ZGWang} Z. G. Wang, \textit{Int. J. of Mod. Phys.} \textit{35}, 2050043 (2020).	
	
     \bibitem{Mao2015}Q. Mao, H­.X. Chen, W. Chen, A. Hosaka, X. Liu and S.L. Zhu,­ {\it Phys. Rev. D} {\bf 92}, 114007 (2015).
	 
	  \bibitem{Chen2015}B. Chen, K.W. Wei and A. Zhang, {\it Eur. Phys. J. A} {\bf 51} 82, (2015).	
	
	  \bibitem{Juhi} J. Oudichchya, K. Gandhi, A. K. Rai, {\it Phys.Rev.D} {\bf 104} 11, 114027, (2021); 103 11, 114030 (2021).
	
	  \bibitem{Wei2017}	K. W. Wei et. al., {\it Phys. Rev. D} {\bf 95}, 116005 (2017).
	
	  \bibitem{Sonnenschein2019}J. Sonnenschein and D. Weissman, {\it Eur. Phys. J. C} {\bf 79}, 326 (2019).
	  
	  \bibitem{Thakkar2017}K. Thakkar, Z. Shah, A.K. Rai and P.C. Vinodkumar, {\it Nucl. Phys. A} {\bf 965}, 57 (2017).

      \bibitem{Roberts2008}W. Roberts and M. Pervin, {\it Int. J. Mod. Phys. A} {\bf 23}, 2817 (2008).	
	   
	  \bibitem{Valcarce} A. Valcarce, H. Garcilazo, J. Vijande, Eur. Phys. J. A 37, 217 (2008).
      
      \bibitem{Ghalenovia2014}Z. Ghalenovi, A. Rajabi, S. x. Qin and D.H. Rischke,  {\it Mod. Phys. Lett. A} {\bf 29}, 1450106  (2014).
   
      \bibitem{Santopinto2005} E. Santopinto, {\it Phys. Rev. C} {\bf 72}, 022201 (2005).
   
      \bibitem{Yoshida} T. Yoshida et. al., {\it Phys. Rev. D} {\bf 92}, 114029 (2015).
    
      \bibitem{Chen2018}K. Chen, Y. Dong, X. Liu, Q.-F. L\"u and T.
     Matsuki, {\it Eur. Phys. J. C} {\bf 78}, 20 (2018).
   \bibitem{Shah2018fbs1}Z. Shah and A.K. Rai, {\it Few-Body Syst.} {\bf 59}, 76 (2018).   
      \bibitem{Padmanath2017} M. Padmanath and N. Mathur, {\it Phys. Rev. Lett.} {\bf 119}, 042001 (2017).
 
    
     \bibitem{Gandhi2018}K. Gandhi, Z. Shah and A. K. Rai, Eur. {\it Phys. J. Plus} {\bf 133}, 512 (2018).
     
     \bibitem{GandhiIJTP2020} K. Gandhi ,Z. Shah, A. K. Rai, Int. J Theor. Phys. 59, 1129–1156 (2020).
    
    \bibitem{AKakadiya} A. Kakadiya et. al., arXiv:2108.11062v1 [hep-ph].
    
    \bibitem{Universe} Z. Shah, A. Kakadiya,  K. Gandhi, A.K. Rai, \textit{Universe} \textbf{7}, 337 (2021).
    
        
     \bibitem{Li2009} B. Q. Li and K. T. Chao, {\it Phys. Rev. D} {\bf 79}, 094004 (2009).
   
   
   	 \bibitem{ShahCPC2016} Z. Shah, K. Thakkar, A. K. Rai, and P. C. Vinodkumar, {\it Chin. Phys. C} {\bf 40}, 123102 (2016).
   	
      \bibitem{Shah2016epja}Z. Shah, K. Thakkar, A.K. Rai and P.C. Vinodkumar, {\it Eur. Phys. J A} {\bf 52}, 313 (2016).
      
      \bibitem{ICC2019} K. Gandhi, A. Kakadiya, Z. Shah, and A. K. Rai,   AIP Conf. Proc. \textbf{2220} (2020), 140015.
      
      \bibitem{DAE2019} A. Kakadiya, K. Gandhi, A. K. Rai,Proceedings of the DAE-BRNS Symp. on Nucl. Phys \textbf{64}, 697 (2019).
      
      \bibitem{Giannini2015} M.M. Giannini and E. Santopinto, {\it Chin. J. Phys.} {\bf 53}, 020301 (2015).
      
      \bibitem{Bijkar2000} R.  Bijkar, F. Iachello, A. Leviatan, {\it Ann. Phys.} {\bf 284}, 89 (2000).
      
     \bibitem{Bijkar1994} R. Bijkar, F. Iachello, A. Laviatan, {\it Ann. Phys. (N. Y.)} {\bf 236}, 69 (1994).
    
     \bibitem{Voloshin2008} M. B. Voloshin , Prog. {\it Part. Nucl. Phys.} {\bf 61}, 455 (2008).
    
     \bibitem{Wang2019} J. Z. Wang, D. Y. Chen,X. Liu, T. Matsuki, {\it Phys. Rev. D} {\bf 99}, 114003 (2019).
    
    \bibitem{Lucha1999} W. Lucha and F. Schoberls, {\it Int. J. Mod. Phys. C} {\bf 10}, 607 (1999).
    \bibitem{Karliner2015}  M. Karliner and J. L. Rosner, {\it Phys. Rev. D} {\bf 92}, 074026 (2015).
    \bibitem{Olive2014} K.A. Olive,  {\it Chin. Phys. C} {\bf 38}, 090001 (2014).
    \bibitem{AaijPRL114} LHCb collaboration, (R. Aaij et. al.), {\it Phys. Rev. Lett.} {\bf 114} 062004 (2015).
    
    \bibitem{AaijPRL2014} LHCb collaboration, (R. Aaij et. al.), {\it Phys. Rev. Lett.} {\bf 113}, 032001 (2014).
    
%
   \bibitem{CMSPRL2012} CMS collaboration, (S. Chatrchyan et al.),	{\it Phys. Rev. Lett.} {\bf 108}, 252002 (2012).
%
\bibitem{AaijPRD2016} LHCb collaboration, (R. Aaij et. al.), {\it Phys. Rev. D} {\bf 93}, 092007 (2016).
     \bibitem{CDF2009} CDF collaboration, T. Aaltonen et al., {\it Phys. Rev. D} {\bf 80},  072003 (2009).
   \bibitem{D02008} D0 collaboration, (V. Abazov et. al.), {\it Phys. Rev. Lett.} {\bf 101}, 232002 (2008).
   
   
  \bibitem{Limphirat}
     A. Limphirat et al., arXiv:0710.3942[hep-ph].
     
     \bibitem{MajethiyaThesis}
  A. MAjethiya,{\it Properties of Heavy Flavour Baryons Using Quark Model} Sardar Patel University, [Thesis].
       
   \bibitem{Majethiya2009}
   A. Majethiya, B. Patel, P. C. Vinodkumar, {\it Eur. Phys. J. A} {\bf 42}, 213 (2009).
    
    \bibitem{Dhir} R. Dhir, C. S. Kim, and R. C. Verma, Phys. Rev. D 88, 094002 (2013).
    
    \bibitem{BPatel} B. Patel, A. K. Rai and P. C. Vinodkumar, \textit{J. Phys. G} \textbf{35} 065001,  (2008) [J. Phys. Conf. Ser. 110 (2008) 122010].
    
    \bibitem{Bernotas} A. Bernotas and V. Simonis, \textit{Lith. J. Phys.} \textbf{53}, 84 (2013).
    
    \bibitem{Franklin}  J. Franklin, D. B. Lichtenberg, W. Namgung, and D. Carydas, \textit{Phys. Rev. D} \textbf{24}, 2910 (1981).
    \bibitem{Pirjol}
     D. Pirjol, T. M. Yan, {\it Phys.Rev. D} {\bf 56}, 5483 (1997).
   \bibitem{ChengPRD2007}
   H. Y. Cheng, C. K. Chua, {\it Phys. Rev. D} {\bf 75}, 014006 (2007).
   
   \bibitem{ChengPRD2017}
   H. Y. Cheng, C. W. Chiang, {\it Phys. rev. D} {\bf 95}, 094018 (2017).
   
   \bibitem{ChengPRD2015}
   H. Y. Cheng and C. K. Chua, {\it Phys. rev. D} {\bf 92}, 074014 (2015). 
   
    \bibitem{dof} Wuensch K L 2011 \textit{Chi-Square Tests} (Berlin: Springer) pp 252–53.
    
    
    \end{thebibliography}
\end{document}